\documentclass[preprint]{elsarticle}
\usepackage[a4paper, top=1in, bottom=1in, left=0.75in, right=0.75in]{geometry}

\usepackage{graphicx}
\usepackage{hyperref}
\usepackage{longtable}
\usepackage{array}
\usepackage{booktabs}
\usepackage{multirow}
\usepackage{url}
\usepackage{hyperref}
\usepackage{tikz}
\usetikzlibrary{arrows.meta} 
\usepackage{forest} 
\usepackage{booktabs}
\usepackage{longtable}
\usepackage{forest}       
\usepackage{tikz}         
\usetikzlibrary{arrows.meta, bending} %
\usepackage{url}
\usepackage{forest} 
\journal{Data in Brief}

\begin{document}

\begin{frontmatter}

\title{Guarding Against Malicious Biased Threats (GAMBiT) Experiments:\\
Revealing Cognitive Bias in Human-Subjects Red-Team Cyber Range Operations}

\author[aff1]{Brandon Beltz}
\author[aff1]{Jim Doty}
\author[aff2]{Yvonne Fonken}
\author[aff3]{Nikolos Gurney}
\author[aff2]{Brett Israelsen}
\author[aff6]{Nathan Lau}
\author[aff5]{Stacy Marsella}
\author[aff1]{Rachelle Thomas}
\author[aff1]{Stoney Trent}
\author[aff2]{Peggy Wu}
\author[aff4]{Ya-Ting Yang}
\author[aff4]{Quanyan Zhu\corref{cor1}}

\cortext[cor1]{Corresponding author}
\ead{qz494@nyu.edu}

\address[aff1]{Bulls Run Group, USA \\ 
\{bbeltz, jdoty, rthomas, stoney\}@bullsrungroup.com}

\address[aff2]{Raytheon Technologies, USA \\ 
\{Brett.Israelsen, Yvonne.M.Fonken, Peggy.Wu\}@rtx.com}

\address[aff3]{Institute for Creative Technologies,  
University of Southern California, USA \\ 
gurney@ict.usc.edu}

\address[aff4]{Department of Electrical and Computer Engineering,  
New York University, New York, NY 10012, USA \\ 
\{yy4348, qz494\}@nyu.edu}

\address[aff5]{Khoury College of Computer Sciences,  
Northeastern University, USA \\
s.marsella@northeastern.edu}

\address[aff6]{Grado Department of Industrial and Systems Engineering,  
Virginia Tech, USA \\
nkclau@vt.edu}

\begin{abstract}
We present three large-scale human-subjects red-team cyber range datasets from the \emph{Guarding Against Malicious Biased Threats (GAMBiT)} project. Across Experiments~1--3 (July~2024--March~2025), 19--20 skilled attackers per experiment conducted two 8-hour days of self-paced operations in a simulated enterprise network (SimSpace Cyber Force Platform) while we captured multi-modal data: self-reports (background, demographics, psychometrics), operational notes, terminal histories, keylogs, network packet captures (PCAP), and NIDS alerts (Suricata). Each participant began from a standardized Kali Linux VM and pursued realistic objectives (e.g., target discovery and data exfiltration) under controlled constraints. Derivative curated logs and labels are included. The combined release supports research on attacker behavior modeling, bias-aware analytics, and method benchmarking. Data are available via IEEE Dataport entries for Experiments 1--3. 
\end{abstract}


\end{frontmatter}

\section*{Specifications Table}

\renewcommand{\arraystretch}{1.2} 
\setlength{\tabcolsep}{4pt}       

\begin{longtable}{>{\raggedright\arraybackslash\bfseries}p{0.28\textwidth} 
                  >{\raggedright\arraybackslash}p{0.68\textwidth}}
\hline
\textbf{Subject} & Computer Networks and Communications; Cybersecurity; Human-Subjects Cyber Operations \\
\hline
\textbf{Specific subject area} & Attacker behavior modeling; intrusion tactics; cognitive biases in cyber operations; network \& host activity data \\
\hline
\textbf{Type of data} & Questionnaires (CSV/JSON); PCAP files; NIDS alerts (Suricata); host keylogs; Bash/Zsh histories; CherryTree operator notes; curated/clean logs \\
\hline
\textbf{How data were acquired} & Live cyber range exercises on the SimSpace Cyber Force Platform with a standardized Kali Linux start box; hourly and end-of-day surveys; host and network sensors (keylogger, shell history, PCAP, Suricata) \\
\hline
\textbf{Data format} & Raw and processed (ZIP archives); derivative curated logs \\
\hline
\textbf{Parameters for data collection} & Two 8-hour sessions per participant; identical starting access; consistent IP addressing per participant range; restricted no-strike subnets; periodic survey prompts \\
\hline
\textbf{Description of data collection} & Self-paced red-team engagements with intermittent intelligence “triggers”; participants documented intended/applied techniques and reasoning; telemetry captured at network choke points and endpoints; synchronized via platform logging \\
\hline
\textbf{Data source location} & United States (EST/EDT) \\
\hline
\textbf{Data accessibility} & IEEE Dataport: \\
& \quad Exp.~1: \url{https://ieee-dataport.org/documents/guarding-against-malicious-biased-threats-gambit-experiment-1} \\
& \quad Exp.~2: \url{https://ieee-dataport.org/documents/guarding-against-malicious-biased-threats-gambit-experiment-2} \\
& \quad Exp.~3: \url{https://ieee-dataport.org/documents/guarding-against-malicious-biased-threats-gambit-experiment-3} \\
& Note: large PCAP files are available only for Experiment 1; PCAP files for other experiments can be provided upon request.\\
\hline
\end{longtable}

\begin{itemize}
\item \textbf{First-of-kind, multi-modal HSR red-team telemetry:}
The dataset provides a uniquely comprehensive capture of human-subject red-team (HSR) exercises in a realistic enterprise cyber range. It synchronizes host-level telemetry (e.g., keystroke logs, shell histories, and operating system event traces), network-level data (full packet captures, intrusion detection system alerts, and NetFlow summaries), and human-factors data (psychometric assessments, targeted surveys, operational notes, and self-reports) collected under standardized, repeatable scenarios. This multi-modal alignment enables fine-grained correlation of attacker actions with cognitive and decision-making processes.
\item \textbf{Bias-aware behavior modeling:}
The experiments are explicitly designed to elicit and measure specific cognitive biases, loss aversion, base rate neglect, availability heuristic, confirmation bias, and sunk-cost fallacy, through carefully designed cyber 'triggers' embedded in the network environment. These triggers, grounded in cognitive science and behavioral economics, provide controlled manipulations to evaluate how biases influence tactical choices, persistence on suboptimal attack paths, and susceptibility to deception. The dataset thus supports the development, training, and validation of bias-sensitive analytics, classifiers, and adaptive defenses.
\item \textbf{Benchmarking realism:}
The experimental environment consists of an enterprise-like network topology with approximately 40 virtual machines per participant range, populated with realistic services, user activity, and operational traffic. Each participant is provisioned with a standardized Kali “start box” for attack execution, along with realistic boundary conditions such as protected “no-strike” network segments. This fidelity allows the data set to serve as a benchmark for evaluating the ecological validity of cyber defense analytics and to compare performance between research teams, algorithms, and operational contexts.
\item \textbf{Reusability and broad applicability:}
Beyond raw telemetry, the dataset includes curated derivatives such as “clean logs,” labeled artifacts, and structured annotations of attacker–trigger encounters. These resources facilitate rapid onboarding into machine learning pipelines, reduce preprocessing burdens, and enable reproducibility studies. The data set is well suited for a range of applications including training and evaluation of intrusion detection systems, development of cognitive vulnerability sensors, educational exercises, and prototyping of adaptive, human-aware cyber defenses.
\end{itemize}

\section{Data Description}

\subsection{Experimental Context and Motivation}

The \emph{Guarding Against Malicious Biased Threats} (GAMBiT) project was established to advance the empirical study of how cognitive biases influence adversary decision-making during realistic, operational-style cyber intrusions \cite{huang2024psyborg+,huang2023cognitive}. Although attacker behavior has been widely examined through theoretical models and simulated environments, there is a notable lack of high-fidelity, human-in-the-loop datasets that capture both detailed operational telemetry and the cognitive states underlying tactical choices. This absence limits the ability of researchers to link observed behaviors directly to the psychological mechanisms that drive them.

GAMBiT addresses this gap through a repeated measures red-team exercise format conducted in a fully instrumented, enterprise-grade cyber range. Within these exercises, controlled manipulations, referred to as \emph{cognitive-bias triggers}, are embedded into realistic attack scenarios. These triggers are designed to induce measurable deviations from optimal or rational strategies and can be systematically attributed to specific heuristics and biases such as loss aversion, base-rate neglect \cite{cox2020stuck}, confirmation and availability bias \cite{shinde2024modeling}, and sunk-cost persistence \cite{pfleeger2012leveraging}. The methodology integrates principles from cyber operations, experimental psychology, and human factors engineering, creating an empirical bridge between cognitive science theory and operational cybersecurity practice \cite{Kamhoua2026AutonomousCyberResilience}.

The experimental design supports two complementary analytic modes. In \emph{within-condition} studies, temporal patterns of bias activation and resolution are examined over the course of a single scenario, revealing how biases emerge, persist and dissipate during ongoing operations. In \emph{cross-condition} comparisons, trigger-present and trigger-absent runs are contrasted to isolate the causal impact of cognitive stimuli on attacker workflows, tool selection, and mission success rates. This dual perspective enables both a detailed characterization of bias manifestation in operational contexts and an assessment of its measurable influence on performance.

From a practical standpoint, the GAMBiT dataset is intended to support the development of bias-aware cyber defense strategies, targeted training interventions, and decision-support systems that account for human cognitive tendencies \cite{chen2018security,huang2020farsighted,hu2023game,Zhu2025CyberDeception,yang2025deceive}. The high temporal and semantic granularity of both behavioral and cognitive data makes it uniquely well-suited for machine learning applications, behavioral modeling, and theory-driven human factors research \cite{kamhoua2021game,nguyen2025bilateral}. Furthermore, the scenario-controlled and repeatable structure of the experiments provides a robust basis for replication, ablation studies, and integration into larger-scale adversarial modeling frameworks.

\subsection{Release Overview}  

The GAMBiT data set comprises three sequential human-subject red-team experiments: HSR1, HSR2, and HSR3 performed in an enterprise-grade cyber range designed for high ecological validity. Each experiment followed tightly controlled, repeatable procedures to enable rigorous measurement of attacker decision-making under varying cognitive-bias conditions. While all phases shared a common network architecture, standardized operational workflow, and synchronized multi-modal data capture framework, they differed in trigger deployment, scenario refinements, and instrumentation depth. These controlled variations were deliberately introduced to distinguish scenario-driven effects from those attributable to cognitive biases, enabling both within-phase and cross-phase behavioral analyses.  

Table~\ref{tab:exp-scope} summarizes the schedule, participant counts, and total data volume for each dataset release.  

\begin{table}[h!]
\centering
\begin{tabular}{l l c c}
\toprule
\textbf{Experiment} & \textbf{Dates (EST/EDT)} & \textbf{Participants} & \textbf{Data Size} \\
\midrule
GAMBiT Exp.~1 (HSR1) & 2024-07-23 -- 2024-09-14 & 19 & 722.8~GB \\
GAMBiT Exp.~2 (HSR2) & 2024-11-09 -- 2025-01-29 & 20 & 2.1~TB \\
GAMBiT Exp.~3 (HSR3) & 2025-02-01 -- 2025-03-26 & 20 & 2.8~TB \\
\bottomrule
\end{tabular}
\caption{Dataset scope by experiment: schedule, cohort size, and archive volume.}
\label{tab:exp-scope}
\end{table}  

\paragraph{Experiment Summaries}  
\begin{itemize}
    \item \textbf{HSR1 (Baseline with Triggers):} Established the foundational scenario and deployed a full suite of cognitive-bias triggers targeting loss aversion, base-rate neglect, availability bias, confirmation bias, and sunk cost persistence. Instrumentation emphasized host and network telemetry, supplemented with pre- and post-session psychometrics. Triggers were embedded along common attack paths to ensure encounter opportunities.  
    \item \textbf{HSR2 (Control Condition):} Removed all bias triggers to capture baseline attacker workflows and natural decision-making patterns. Human-subjects instrumentation was expanded to include hourly reasoning and affect surveys, enabling finer alignment between cognitive state and operational behavior. Network capture coverage was extended to additional vantage points, increasing the dataset’s scope relative to HSR1.  
    \item \textbf{HSR3 (Refined Triggers):} Reintroduced bias triggers in a more targeted and context-specific manner, informed by findings from HSR1 and HSR2. Refinements included timestamped encounter labels, co-occurring bias codes, and optimized placement to maximize encounter rates. Instrumentation upgrades featured enhanced keystroke sessionization, expanded Suricata rule sets, and additional Expert Knowledge Model (EKM) sensors for near-real-time classification of ``rational'' versus ``biased'' actions.  
\end{itemize}  

Across all three phases, data capture was synchronized across modalities (network, host, and survey) with standardized file formats, metadata conventions, and consistent naming schemes to support automated parsing and robust cross-phase comparative analytics.  


\section{Data Access and Structure}

\subsection{Source and Accessibility}
Data were collected in the United States and time-stamped in Eastern Time (EST/EDT). All releases are hosted on IEEE DataPort with persistent links:
\begin{longtable}{@{}ll@{}}
\toprule
Data source location & United States (EST/EDT) \\
Data accessibility   & IEEE DataPort: \\
& \quad Exp.~1: \url{https://ieee-dataport.org/documents/guarding-against-malicious-biased-threats-gambit-experiment-1} \\
& \quad Exp.~2: \url{https://ieee-dataport.org/documents/guarding-against-malicious-biased-threats-gambit-experiment-2} \\
& \quad Exp.~3: \url{https://ieee-dataport.org/documents/guarding-against-malicious-biased-threats-gambit-experiment-3} \\
\bottomrule
\end{longtable}

\subsection{Top-Level Organization}
Each experiment (HSR1, HSR2, HSR3) is distributed as a compressed archive (\texttt{.zip} or \texttt{.tar.gz}). At the archive root:
(i) a set of participant-indexed ZIP bundles, one per participant, and
(ii) experiment-level documentation (README, schemas, checksums, notes).
Participant bundles are named with zero-padded numeric IDs, e.g., \texttt{P01.zip}, \texttt{P02.zip}, \dots, \texttt{PXX.zip}, preserving lexicographic order and simplifying programmatic access.
There is also a cleaned-up spreadsheet (\texttt{.xlsx}) that consists of the responses/scores for the screening, demographics, psychometrics, and Kali-command-to-MITRE-techniques for all experiment participants.

\subsection{Per-Participant Contents (Modality Folders)}
Within each participant ZIP, data are organized by modality in a consistent hierarchy:
\begin{itemize}
  \item \textbf{\texttt{self\_reports/}} (or \texttt{screening\_demographics/}): baseline screening \& demographics in CSV/XLSX.
  \item \textbf{\texttt{psychometrics/}}: CRT, BFI-2-XS, GRiPS, A-DMC item-level responses and scored sheets (CSV/XLSX).
  \item \textbf{\texttt{questionnaires/}}: hourly and end-of-day reports (intended vs.\ applied ATT\&CK, reasoning/affect in HSR2--HSR3), CSV/XLSX.
  \item \textbf{\texttt{opnotes/}}: CherryTree operational notes (XML/CTD).
  \item \textbf{\texttt{network/}}: PCAP captures and Suricata \texttt{eve.json}; may include summary tables (XLSX/CSV).
  \item \textbf{\texttt{host/}}: keylogger streams, clipboard text, and timestamped \texttt{.bash\_history}/\texttt{.zsh\_history}.
  \item \textbf{\texttt{derived/}}: curated ``clean logs'', trigger-encounter annotations, and multi-modal alignment tables (CSV/XLSX).
\end{itemize}

\subsection{File Types and Documentation}
Raw telemetry is provided as \texttt{.pcap}, \texttt{.json}, \texttt{.log}, and CherryTree XML; tabular summaries and schemas as \texttt{.csv}/\texttt{.xlsx}; notes and guidance as \texttt{.md}/\texttt{.txt}. Each experiment includes:
\begin{itemize}
  \item \textbf{README} (archive overview, usage notes),
  \item \textbf{schemas} (variable/field definitions per modality),
  \item \textbf{checksums} (SHA-256) for integrity verification,
  \item \textbf{experiment notes} (scenario configuration, trigger status, known anomalies).
\end{itemize}

\subsection{Indexing and Naming Conventions}
Participants are indexed numerically (\texttt{P01}, \texttt{P02}, \dots), and file names embed key descriptors to support automated fusion across modalities. A typical pattern is:
\[
\texttt{<ExpID>-\!<ParticipantID>-\!<Modality>-\!<YYYYMMDD\_HHMMSS>.<ext>}
\]
e.g.,
\texttt{HSR2-P07-NET-20250112\_143005.pcap},
\texttt{HSR2-P07-HOST-20250112\_143010.json},
\texttt{HSR2-P07-QUESTIONNAIRES-20250112\_EOD.xlsx}.
Timestamps are synchronized across modalities using the range time source (EST/EDT), enabling direct temporal alignment without additional preprocessing.

\subsection{Directory Diagram}
\begin{figure}[h!]
\centering
\begin{forest}
for tree={
  font=\ttfamily\small,
  grow'=0,                
  parent anchor=east,     
  child anchor=west,      
  anchor=west,
  align=left,
  inner sep=2pt,
  minimum width=4.0cm,    
  l sep=20pt,             
  s sep=10pt,             
  edge path={%
    \noexpand\path[draw, >={Stealth[round]}, ->]
      (!u.east) -- (.west);}, 
}
[HSR2\_archive.zip
  [README.md]
  [schemas.xlsx]
  [checksums.sha256]
  [P01.zip
    [self\_reports/
      [demographics\_P01.xlsx]
    ]
    [psychometrics/
      [CRT\_P01.xlsx]
      [BFI2XS\_P01.xlsx]
      [GRiPS\_P01.xlsx]
      [ADMC\_P01.xlsx]
    ]
    [questionnaires/
      [hourly\_20250112\_P01.xlsx]
      [eod\_20250112\_P01.xlsx]
    ]
    [opnotes/
      [opnotes\_P01.ctd]
    ]
    [network/
      [HSR2\_P01\_20250112\_143005.pcap]
      [HSR2\_P01\_suricata\_eve.json]
      [flows\_summary\_P01.xlsx]
    ]
    [host/
      [HSR2\_P01\_keylog\_20250112.json]
      [bash\_history\_P01.txt]
      [zsh\_history\_P01.txt]
    ]
    [derived/
      [clean\_logs\_P01.csv]
      [trigger\_encounters\_P01.csv]
      [alignment\_timeline\_P01.xlsx]
    ]
  ]
  [P02.zip]
  [{\ldots}]
]
\end{forest}
\caption{Directory layout: experiment-level files and participant-indexed bundles; each participant folder contains modality-specific subdirectories and representative raw/derived files.}
\label{fig:dir-tree}
\end{figure}
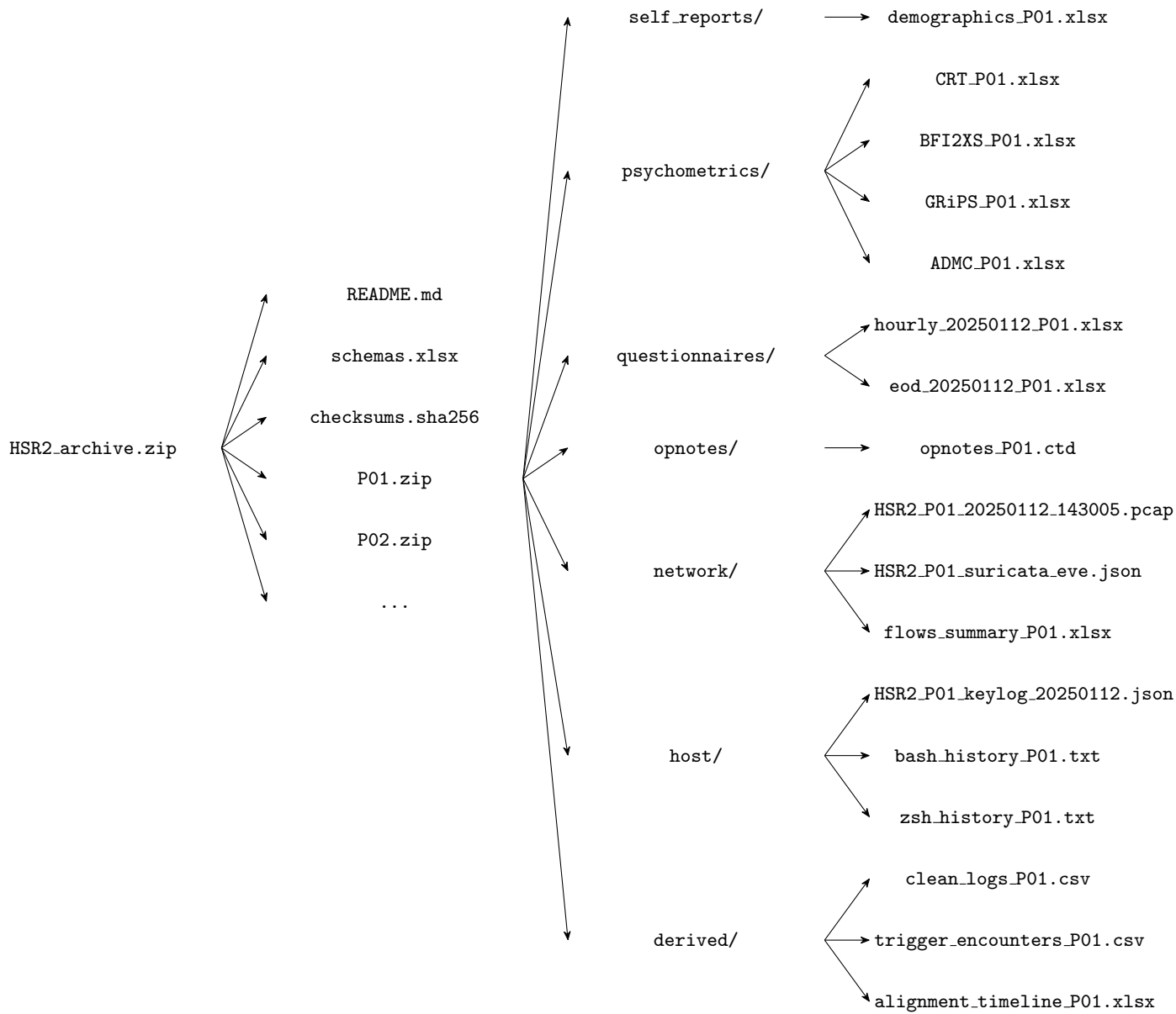

Figure~\ref{fig:dir-tree} illustrates the experiment $\rightarrow$ participant $\rightarrow$ modality hierarchy and representative files.

 

The release includes raw, processed, and derivative data across six primary modalities, organized in a standardized directory structure with SHA-256 checksums for integrity verification:

\begin{enumerate}
    \item \textbf{Self-reports:} Baseline screening and demographic forms; psychometric batteries including the Cognitive Reflection Test (CRT), Big Five Inventory-2 Short/Extra-Short Forms (BFI-2-S / BFI-2-XS), General Risk Propensity Scale (GRiPS), and Adult Decision-Making Competence (A-DMC) subscales. Administered pre-scenario to characterize participant profiles.
    \item \textbf{Questionnaires:} Hourly and end-of-day forms capturing intended vs.\ applied MITRE ATT\&CK techniques, operational reasoning, and affect/motivation measures (the latter added in HSR2--HSR3). These link subjective reasoning to objective behavioral traces.
    \item \textbf{Operational Notes (OPNOTES):} Participant-generated logs in CherryTree format, documenting perceived goals, network observations, and tactical decisions.
    \item \textbf{Network data:} Full packet captures (PCAP) from all subnets and Suricata intrusion detection alerts, aligned with the network clock and tagged per participant session.
    \item \textbf{Kali host data:} Continuous keylogger streams and shell histories (\texttt{.bash\_history}, \texttt{.zsh\_history}), enabling reconstruction of exact command sequences with timing.
    \item \textbf{Derivative products:} Curated ``clean logs'' produced via automated and manual processing to improve readability and remove sensitive range management artifacts, suitable for direct use in analytics pipelines.
\end{enumerate}

The repository metadata describes the complete directory hierarchy, file naming conventions, and modality-specific schema definitions.

\section{Experimental Design, Materials, and Methods}

\subsection{Cyber Range Architecture and Experimental Setup}

The GAMBiT cyber range was provisioned on the SimSpace Cyber Force Platform and configured to emulate the operational environment of a mid-sized enterprise. The virtualized infrastructure comprised approximately 40 endpoints, including both server and workstation roles, running representative business services such as web hosting, file sharing, email, and directory services. Interconnection was achieved through virtual routers and switches configured with realistic internal subnetting and routing policies. To sustain an authentic background environment, synthetic user traffic, including web browsing, email exchanges, and file transfers, was continuously generated, camouflaging participant activity within normal operational patterns.

Each participant operated within a logically and physically isolated clone of a common baseline configuration. This isolation preserved experimental control, ensured repeatability, and prevented cross-contamination of telemetry across participants. The network replicated enterprise defense-in-depth architectures, including intrusion detection systems (IDS) and segmented security zones, to reinforce realism and operational complexity.

The standardized starting point for all red-team operations was a Kali Linux virtual machine (\texttt{10.10.0.5}) provisioned with a curated toolkit for reconnaissance, exploitation, and post-exploitation. This uniform capability baseline allowed skill-level differences to emerge organically in operational traces.

To safeguard range management infrastructure and maintain exercise fidelity, several \emph{no-strike} network segments were designated and enforced:
\[
\texttt{10.10.0.0/16}, \quad \texttt{155.41.3.0/24}, \quad \texttt{192.168.0.0/21}, \quad \texttt{172.16.100.0/22}, \quad \texttt{3.136.223.108}.
\]
Participants were instructed not to scan or interact with these segments, and violations were automatically detected and logged through firewall rules and monitoring agents.

The dataset release includes complete topology diagrams, IP allocation tables, and per-host service inventories. These resources detail OS distributions, active services, routing architectures, and simulated user account structures, enabling precise contextualization of captured network and host data.

\begin{figure}[h!]
\centering
\includegraphics[width=0.7\linewidth]{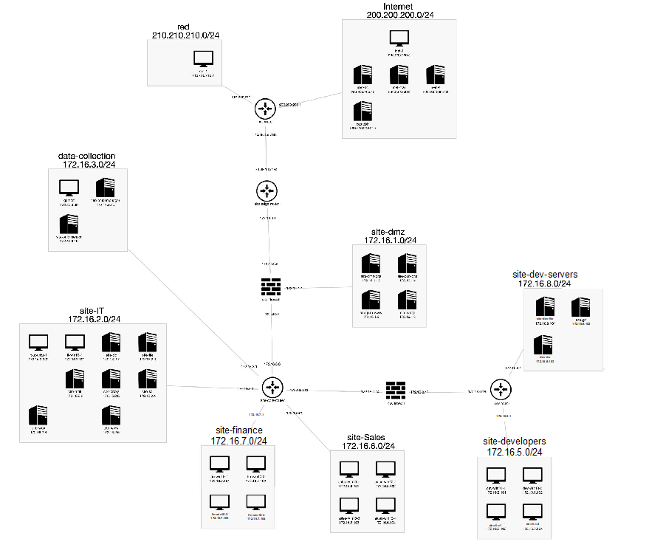} 
\caption{Representative GAMBiT cyber range topology showing core services, segmented zones, and participant start node.}
\label{fig:topology}
\end{figure}

\subsection{Bias Triggers and Labeling Protocols}

Bias triggers were embedded directly into the cyber range scenarios as controlled manipulations, jointly designed by cybersecurity SMEs and cognitive scientists to induce bias-prone decision-making under operationally realistic. Each trigger was mapped to a primary cognitive bias, linked to relevant MITRE ATT\&CK tactics and techniques, and implemented in the network as a plausible file, credential, service, or configuration. The design process emphasized both technical plausibility and psychological efficacy, with iterative testing to ensure that each trigger could be encountered naturally along common attack paths. Representative examples include:
\begin{itemize}
    \item \emph{Loss Aversion:} A notes file referencing potentially valuable credentials that offered faster access but carried elevated detection risk, contrasted with a slower but safer exploitation.
    \item \emph{Availability Heuristic:} Directories seeded with memorably named yet strategically irrelevant files, diverting attention from critical assets.
    \item \emph{Base-Rate Neglect:} User accounts configured with misleading administrative naming conventions, encouraging unverified privilege assumptions.
    \item \emph{Sunk Cost / Confirmation Bias:} Deceptive artifacts or aliased commands that incentivized persistence along unproductive or dead-end attack paths.
\end{itemize}

From Experiment 1 to Experiment 2, trigger placement was refined to increase encounter rates by positioning them in higher-traffic network locations and integrating supplementary logging sources for richer post-event. All triggers were labeled with a structured code indicating bias type, trigger class, and instance number (e.g., \texttt{B.2.1.1} for a Base-Rate Neglect trigger of class~2, instance~1) to support rapid identification and consistent analysis.

In HSR2 and HSR3, the labeling framework was enhanced to include:  
(i)~precise timestamps for each trigger encounter,  
(ii)~standardized bias codes, and  
(iii)~annotations from the Expert Knowledge Model (EKM) sensors indicating whether subsequent actions aligned with ``rational'' or ``biased'' decision.  

These refinements enabled fine-grained statistical and modeling analyses of how specific cognitive biases influenced adversary decision-making, revealing measurable impacts on attacker progress, resource expenditure, and operational detectability in complex, time-constrained operations.

In HSR2 and HSR3, the labeling framework was enhanced to include precise trigger encounter timestamps, associated bias codes, and annotations indicating whether observed actions aligned with the ``rational'' or ``biased'' paths defined by the EKM sensor suite. These refinements enable fine-grained statistical and modeling analyses of how cognitive biases shape adversary decision-making in complex, time-constrained operations.

\subsection{Scenario and Objectives}
Each experiment placed participants in a realistic, enterprise-scale cyber range simulating a mid-sized organization with multiple subnets, operational services, and user traffic. Participants were given initial access from a standardized Kali Linux ``start box'' and a briefing outlining broad operational objectives: infiltrate the network, expand access, map services, identify high-value assets, and exfiltrate sensitive files.  

The scenario was intentionally open-ended, allowing participants to self-direct their tactics, techniques, and procedures (TTPs) over the course of a multi-stage, multi-hour engagement. Across the experiments, objectives were periodically updated every four hours to simulate evolving mission priorities (e.g., collect system administrator details, enumerate developer network, or acquire finance records).  

Participants recorded their thought processes, decision points, and operational steps through:
\begin{enumerate}
    \item \emph{Hourly reports} – detailing intended and executed MITRE ATT\&CK techniques, as well as responses to targeted reasoning and affect questions (in Exp.~2--3).
    \item \emph{Structured operational notes} (CherryTree format) – providing narrative context, network mapping results, and intermediate findings.
    \item \emph{End-of-day summaries} – reflecting on successes, challenges, engagement level, and perceived adaptation.
\end{enumerate}

This design enabled direct linkage of subjective reasoning and affect with objective behavioral telemetry, facilitating both cognitive and operational analysis.

\subsection{Environment and Instrumentation}
Experiments were conducted using the SimSpace Cyber Force Platform, which provisioned identical, isolated ranges for each participant. The environment was preconfigured with $\sim$40 virtual hosts spanning finance, sales, developer, and DMZ segments, with realistic business services and synthetic user traffic.  

Instrumentation spanned three primary layers:
\begin{itemize}
    \item \textbf{Network-level capture:}  
    Full packet capture (PCAP) at range egress points; Suricata IDS rulesets tuned to detect suspicious or policy-violating traffic, generating time-stamped alerts for correlation with host and survey data.
    \item \textbf{Host-level capture (Kali start box):}  
    Continuous keylogging, including keystrokes and clipboard activity, to capture reconnaissance, exploitation, and post-exploitation commands.  
    Timestamped Bash and Zsh command histories to support command sequence reconstruction.
    \item \textbf{Human-subjects instruments:}  
    Pre-scenario psychometric batteries — Cognitive Reflection Test (CRT), Big Five Inventory-2 Extra-Short Form (BFI-2-XS), General Risk Propensity Scale (GRiPS), and Adult Decision-Making Competence (A-DMC) subscales.  
    Hourly and end-of-day surveys capturing tactics used, reasoning, affective state, and trigger recognition (Exp.~2--3 added Likert-scale reasoning/affect items).
\end{itemize}

All telemetry streams were clock-synchronized to enable cross-modal correlation.

\subsection{Derived Artifacts}
In addition to raw captures, the dataset includes derivative products to support rapid analysis:
\begin{itemize}
    \item \emph{Curated ``clean logs''} – keylogger streams processed to remove extraneous or repeated keystrokes while preserving operational context and temporal fidelity.
    \item \emph{Multi-modal alignment files} – per-participant, per-session aggregates linking network events, host commands, IDS alerts, and survey responses into a unified timeline.
    \item \emph{Trigger encounter annotations} – labeling instances where participants interacted with bias-targeted artifacts, noting associated MITRE ATT\&CK techniques and Expert Knowledge Model (EKM) bias sensor outputs.
\end{itemize}

\subsection{Hypotheses and Bias Constructs}

The GAMBiT experimental framework was designed to evaluate two central hypotheses.  
First, attacker proficiency level, operationalized as division assignment (\emph{open} vs.\ \emph{expert}), predicts the rate of progress along a predefined ``ideal'' attack path, which represents the most efficient sequence of steps leading to scenario objectives.  
Second, the deliberate introduction of cognitive-bias triggers into the operational environment produces measurable deviations in attacker decision-making, resource allocation, and tactical sequencing.

Bias constructs were instantiated in scenario design through specific, controlled stimuli intended to elicit well-documented cognitive heuristics and systematic errors. These included:

\begin{itemize}
    \item \textbf{Loss aversion:} The tendency to overvalue the preservation of existing gains, such as persisting with a risky shortcut (e.g., planted credentials) rather than transitioning to a slower but more reliable exploitation method.
    \item \textbf{Base-rate neglect:} Disregarding prior probability distributions, for example assuming that accounts labeled with ``-adm'' automatically possess elevated privileges without verification.
    \item \textbf{Availability heuristic:} Overweighting salient or recent information, such as prioritizing files with conspicuous names regardless of actual content relevance.
    \item \textbf{Confirmation bias:} Selectively seeking evidence that reinforces an existing working hypothesis, for instance repeatedly attempting to use invalid SSH keys while rationalizing authentication failures.
    \item \textbf{Sunk cost fallacy:} Persisting with ineffective tactics due to previous investment of time or effort, such as continuously targeting a protected file of minimal strategic value.
\end{itemize}

By combining these targeted bias-inducing manipulations with high-fidelity operational data and synchronized human-subject measurements, GAMBiT enables rigorous empirical analysis of attacker cognition, decision-making under uncertainty, and susceptibility to defensive deception \cite{pawlick2021game}. This design supports both statistical inference and computational modeling of bias dynamics in adversarial contexts, offering an empirical basis for bias-aware defensive strategies.

\section{Usage Notes}

\subsection{Access and Organization}
The GAMBiT dataset is distributed as three experiment-specific archives (HSR1, HSR2, HSR3), each hosted on IEEE Dataport with persistent DOI links provided in the \emph{Specifications Table}. This segmentation allows users to selectively download only the phases relevant to their analysis or to work across all three for comparative studies. 

Each archive is packaged as a compressed file (\texttt{.zip} or \texttt{.tar.gz}) and follows a consistent internal directory hierarchy to simplify navigation and automated ingestion. At the top level, the archive contains:

\begin{itemize}
    \item \textbf{Modality-specific subdirectories}, organized by data source type, such as \texttt{network/} for packet captures and intrusion detection alerts, \texttt{kali\_host/} for host-level activity logs, \texttt{surveys/} for self-report and psychometric data, \texttt{opnotes/} for participant-generated operational notes, and \texttt{derived/} for cleaned or aggregated files. This separation ensures that researchers can focus on a single modality without unnecessary file handling.
    \item \textbf{Metadata files} providing essential context for analysis. These include \texttt{README} documents with an overview of data contents, modality-specific schema definitions describing variable names and formats, checksum lists (SHA-256) for verifying data integrity, and experiment-level notes detailing any scenario-specific conditions or anomalies.
    \item \textbf{Derived artifacts}, which are preprocessed, analyst-ready resources such as curated ``clean logs'' stripped of non-essential range management data, trigger encounter annotations marking cognitive-bias stimuli, and multi-modal alignment files linking events across different telemetry sources via synchronized timestamps.
\end{itemize}

This organizational structure accommodates a wide range of workflows. Researchers can work directly with the raw data for custom processing, reproducibility, and forensic reconstruction, or they can load the derived products to accelerate prototyping, machine learning experiments, and exploratory analyses without investing in extensive preprocessing.

\subsection{Recommended Tools}
Due to its multi-modal nature, the dataset can be approached with a variety of analytical toolchains, depending on the user's domain focus and methodological expertise. The following recommendations are not exhaustive but represent common starting points for working with GAMBiT data:

\begin{itemize}
    \item \textbf{Network captures (PCAP) and IDS alerts:} Tools such as \texttt{tcpdump} can be used for targeted packet filtering and extraction, while \texttt{tshark} or Wireshark provide interactive packet inspection with protocol-level decoding. Suricata’s \texttt{eve.json} output can be parsed using Python or Logstash pipelines for signature-based intrusion event analysis and correlation with other modalities.
    \item \textbf{Host-level logs (keystrokes, shell histories):} Standard text-processing utilities like \texttt{grep}, \texttt{ripgrep}, or \texttt{awk} are useful for rapid keyword searches. For structured log parsing and temporal analysis, JSON processors like \texttt{jq} and data analysis libraries in Python (\texttt{pandas}) or R (\texttt{dplyr}, \texttt{tidyverse}) enable session segmentation, command sequence reconstruction, and feature extraction for statistical or machine learning models.
    \item \textbf{Survey and psychometric data:} The structured CSV and JSON formats are compatible with statistical packages in Python (\texttt{pandas}, \texttt{numpy}, \texttt{scipy}, \texttt{statsmodels}) and R (\texttt{psych}, \texttt{tidyverse}). These can be used to score psychometric scales, compute internal consistency measures (e.g., Cronbach’s alpha), perform inferential tests, or integrate self-report measures with operational telemetry for mixed-methods analysis.
\end{itemize}

To streamline multi-modal analysis, all data streams have been normalized to a consistent timestamp format and synchronized to the range’s NTP service. This enables straightforward cross-referencing of events between network, host, and self-report modalities without the need for substantial temporal alignment preprocessing. Where possible, file naming conventions embed participant identifiers, modality codes, and session timestamps, further reducing the overhead in linking disparate data sources.

\subsection{Use Cases}

There are several use cases for this dataset, spanning cybersecurity research, machine learning, human factors analysis, education, and strategic policy work. The combination of multi-modal telemetry, human factors data, and controlled yet realistic enterprise-like scenarios enables both methodological development and empirical validation.

\subsubsection{Cybersecurity Research and Analytics}
The dataset supports investigations into attacker behavior modeling, adversarial decision-making, and cognitive bias analysis. By linking operational telemetry with survey-based self-reports, researchers can study how decision-making evolves under bias-triggering conditions, examine the relationship between attacker skill indicators and operational outcomes, and compare the effectiveness of heuristic versus machine learning approaches in detecting behavioral signatures of bias. These capabilities make the dataset well-suited for work in adversarial reasoning, deception detection, and behavior-based intrusion detection \cite{li2024symbiotic,zhu2025game,li2025texts}.

\subsubsection{Machine Learning and AI Benchmarking}
 curated ``clean logs'' and labeled annotations enable reproducible benchmarks for predictive models, such as sequence classification to anticipate the next command or tactic, bias detection models that integrate host, network and psychometric features, and cross-modal fusion architectures that combine text, time series and network data. Because the scenarios capture realistic and varied attacker workflows, they are ideal for assessing the robustness and generalizability of machine learning pipelines.

\subsubsection{Human Factors and Cognitive Security Studies}
The inclusion of psychometric and affective measures offers a rare opportunity to link cognitive traits and decision-making processes to operational performance. Researchers can explore how individual differences, such as cognitive reflection or risk perception \cite{frederick2005cognitive}, influence cyber operations, compare self-reported reasoning to observed actions, and assess correlations between bias susceptibility and task efficiency or success. This supports studies in human-in-the-loop security, cognitive load modeling, and attacker persona profiling.

\subsubsection{Education and Training}
Because the environment is realistic yet fully controlled, the dataset serves as a resource for cybersecurity curriculum development, where students can analyze authentic adversary behavior without direct system access. It also supports red team/blue team training exercises and digital forensics practice, using PCAP and host telemetry for event reconstruction. Replayable scenarios and annotated datasets allow learners to experiment with diverse analytical approaches.

\subsubsection{Policy and Strategy Development}
Beyond technical and educational applications, the dataset informs policy-oriented research by providing empirical evidence for bias-aware defensive measures, enabling evaluation of training programs designed to reduce decision-making errors, and supporting wargaming scenarios that incorporate realistic attacker cognition models \cite{yang2025multi,huang2024psyborg+}. Its structure, annotations, and multi-modal scope make it valuable to both practitioners and decision-makers seeking data-driven insights.


\section{Related Datasets}
\label{sec:related-datasets}

Several publicly available datasets capture human behavior and decision-making in cybersecurity contexts, with explicit or implicit connections to cognitive bias. These resources provide valuable baselines and complementary data for studying the mechanisms underlying biased judgment, susceptibility to manipulation, and operational decision-making under uncertainty.

\paragraph{PsyScam Benchmark}
The PsyScam dataset \cite{ma2025psyscam} is a curated collection of online scam communications, sourced from multiple public reporting platforms, and annotated according to established psychological and cognitive influence techniques. It supports supervised learning tasks such as tactic classification, scam type identification, and automated generation of realistic deceptive content. By grounding annotations in cognitive science theory, PsyScam enables systematic study of persuasion mechanisms relevant to social engineering and phishing.

\paragraph{Cry Wolf Dataset}
The Cry Wolf dataset \cite{roden2020crywolf} investigates the phenomenon of \emph{alert fatigue} in cybersecurity operations, focusing on analyst responses to intrusion detection system (IDS) alerts, particularly ``impossible travel'' events. It contains both alert metadata and analyst classification outcomes (true positive vs. false positive), making it suitable for analyzing decision-making degradation in noisy environments. While not explicitly bias-annotated, the dataset aligns with research on cognitive overload and trust calibration in human--machine teaming.

\paragraph{ReSCIND PsyCCDEF Dataset}
Developed under the IARPA ReSCIND program, the PsyCCDEF dataset \cite{gonzalez2025psyccdef} embeds classic cognitive bias elicitation tasks within ecologically valid cyber problem-solving scenarios. Participant responses, decision times, and accuracy are recorded, demonstrating that 81\% of observed biases from the traditional literature also manifest in cyber-relevant settings, often with measurable performance costs. This dataset offers a rare opportunity to examine bias persistence in domain-specific operational contexts.

\paragraph{Comparison with the HSR Dataset}
While the above datasets each contribute valuable insights, the \emph{HSR} dataset differs in several key aspects. First, it combines \textbf{multi-modal raw telemetry} (e.g., network packet captures, host-level logs, keystroke data) with \textbf{fine-grained psychometric assessments} administered longitudinally across experimental sessions, enabling cross-temporal modeling of cognitive bias dynamics. Second, unlike the CTF and Cry Wolf datasets, which focus on narrowly scoped operational tasks, HSR encompasses a \textbf{broad task ecology} including self-report surveys, psychometrics, scripted cyber exercises, and naturalistic behavioral interactions. Third, compared to PsyScam, which contains only attacker-generated content, HSR records \textbf{both stimuli and participant behavioral responses}, making it suitable for end-to-end modeling of bias influence and mitigation. Finally, while PsyCCDEF places bias measurement within cyber scenarios, it does not include rich raw data streams for replay or forensic analysis; HSR's inclusion of such streams facilitates \textbf{multi-level analysis} spanning cognitive, behavioral, and system-interaction layers.

\section{Limitations}
No-strike segments constrain some reconnaissance behaviors; range artifacts (\emph{triggers}) may shape action sequences; keylogging and shell histories reflect the Kali jump host (other hosts instrumented via network vantage). Survey compliance/timing may vary by participant.

\section*{Ethics Statement}
Data were collected as human-subjects research (HSR) with informed procedures appropriate to the program and context. Releases include operational notes and survey responses from consenting participants, with data curated for research access via controlled repositories. Users should adhere to the repository access terms and any downstream IRB requirements.

\section*{Declaration of Competing Interest}
The authors declare no competing interests.

\section*{Acknowledgments}
We acknowledge the GAMBiT/ReSCIND performer team and platform support, and contributors to scenario design, execution, curation, and analysis.

\section*{Supplementary Materials}
Topology diagrams and host inventories; survey instruments and scoring directions; example parsing scripts; data schemas and dictionaries.

\bibliographystyle{plain} 
\bibliography{references} 

\end{document}